# Geometric Asymmetry-Enhanced Nonreciprocal Supercurrent Transport Revealed by Second-Harmonic Response

Yu He[1], Zifeng Wang[1], Jiaxu Li[1], Fenglin Zhong[2], Haozhe Yang[1], Kewen Shi[1], Le Wang[3†], Guang Yang[1‡], Weisheng Zhao[1]

[1]*Fert Beijing Institute, MIIT Key Laboratory of Spintronics, School of Integrated Circuit Science and Engineering, Beihang University, Beijing 100191, China*

[2]*Darwin College, University of Cambridge, Cambridge CB3 9EU, UK*

[3]*Beijing Key Laboratory of Optoelectronic Functional Materials and Micro-nano Devices, School of Physics, Renmin University of China, Beijing 100872, China*

[†]Corresponding author: le.wang@ruc.edu.cn (Le Wang)

[‡]Corresponding author: gy251@buaa.edu.cn (Guang Yang)

Nonreciprocal transport in superconducting systems serves as a powerful probe of symmetry-breaking mechanisms, with the superconducting diode effect emerging as a key manifestation enabling cryogenic rectification. While theoretical models have extensively explored superconducting nonreciprocity, experimental verification remains challenging, as conventional transport measurements struggle to disentangle intrinsic and extrinsic contributions. Nonlinear transport analysis, particularly second-harmonic response, offers an alternative approach by providing a sensitive probe for detecting spatial inversion symmetry breaking in the presence of time-reversal symmetry violation. Here, we systematically investigate the influence of geometric symmetry on nonreciprocal transport by comparing two triangular-extended Hall bar configurations with a symmetric Hall bar control. Second-harmonic nonlinear transport measurements reveal that the triangular extension significantly enhances nonreciprocal response, exhibiting a clear dependence on the base angle of the extension. These findings establish a direct connection between mesoscopic geometry and macroscopic nonreciprocity, demonstrating how spatial symmetry and vortex dynamics govern nonlinear transport. This insight offers a guiding principle for designing superconducting rectification architectures.

# I. INTRODUCTION

The diode effect, a quintessential example of nonreciprocal transport, is a fundamental component of modern semiconductor circuits. A prominent realization is the p-n junction diode, where structural inversion symmetry breaking enables rectification, allowing preferential forward conduction while suppressing reverse current in the current-voltage (*I*-*V*) characteristics. Extending this concept to superconducting systems has led to the emergence of the superconducting diode effect (SDE), which holds significant promise for dissipationless rectifying devices[1–3]. In such systems, supercurrent flows without dissipation in one direction, whereas in the opposite direction, charge transport occurs via normal conducting electrons. This breakthrough has the potential to enable energy-efficient quantum technologies and advance applications in quantum information processing [2–5].

Unlike conventional charge carrier transport, which is independent of time-reversal symmetry, supercurrents—persisting in the absence of an external voltage—are inherently sensitive to time-reversal operations. This sensitivity necessitates the simultaneous breaking of both spatial inversion and time-reversal symmetries for nonreciprocal supercurrent transport[6–8]. Rikken extended the study of polarization-independent optical effects to magneto-transport in non-centrosymmetric conductors, establishing the theoretical framework of magnetochiral anisotropy (MChA) to describe nonlinear response characteristics[6,9]:

$$R = \mathrm{R}_0(1 + \gamma BI) \qquad \textbf{(Eq. 1)}$$

where $\gamma$ quantifies the magnitude of MChA, a phenomenon exclusively observed in non-centrosymmetric crystalline systems.

Recent theoretical and experimental studies have identified two primary mechanisms underlying the SDE: (i) intrinsic mechanisms, including spin-orbit coupling with Zeeman fields[10–16] or finite-momentum Cooper pairs[17–19], and (ii) extrinsic factors, such as geometric asymmetries (e.g., vortex pinning[20–22], nano-structuring[22–26]) and interfacial effects at heterojunctions[12,19,27–29]. While Hall bar devices ideally retain centrosymmetry, the long-range phase coherence of superconductors amplifies even minor defects in symmetric superconducting strips[13,19,30–32] and unavoidable edge roughness from nanofabrication[19,33,34], significantly enhancing symmetry breaking effects. A perpendicular magnetic field component further modifies the surface potential barrier, facilitating the entry of Abrikosov vortices carrying magnetic flux quanta into the superconductor[33,35–39]. Initially, these vortices pin at intrinsic defect sites to minimize the system's free energy. Upon application of an electric current, the vortices experience a Lorentz force that drives them perpendicular to the current flow. The motion of vortex-antivortex pairs induces localized electric fields within the superconductor, disrupting the characteristic zero-resistance state and leading to energy dissipation via Joule heating. Prior studies have demonstrated that, even in zero-field conditions, thermally and electrically activated vortices can exhibit a Hall-like effect, accumulating at opposing channel boundaries due to finite temperature gradients and current-induced dynamics[40]. Consequently, vortex dynamics exhibit directional selectivity, where asymmetric boundary conditions and the Meissner screening effect govern the preferential motion of vortices and antivortices toward opposite channel edges[19,25,33,34,41,42].

While conventional transport measurements provide insight into nonreciprocal effects, they often struggle to disentangle intrinsic and extrinsic contributions. In contrast, second-harmonic nonlinear transport measurements, particularly harmonic response analysis, serve as a powerful tool for detecting symmetry-breaking mechanisms with high sensitivity. By probing the second-harmonic voltage response, this approach enables a more direct quantification of spatial inversion symmetry breaking under time-reversal symmetry violation[7,8,10,12,32,43–46]. In this work, we systematically investigate the influence of geometric symmetry on nonreciprocal transport by comparing three Nb devices with and without triangular extensions[25]. Analysis of *I-V* characteristics reveals distinct stages in vortex dynamic during the superconducting-to-normal transition, including vortex generation, creep, flow, and annihilation[47–51]. Furthermore, *I-V* measurements under varying external magnetic fields demonstrate pronounced transport nonreciprocity within a narrow field range, quantified by the directional-dependent critical current ($I_{c,\pm}$) and the quality factor $Q = \frac{2(I_{c,+} - |I_{c,-}|)}{I_{c,+} + |I_{c,-}|}$ reaching values of up to around 20%[52]. The nonreciprocal magnetochiral anisotropy coefficient $\gamma$ is further extracted from harmonic measurements[6]. Notably, $\gamma$ exhibits a distinct peak under varying temperatures and bias currents (while remaining below the critical threshold), strongly correlating with features observed in *I-V* analysis. Through a systematic investigation, we establish that variations in $\gamma$ across devices primarily arise from the angular relationship between the extension direction and the channel edge orientation. This geometric configuration plays a crucial role in modulating vortex density and velocity, ultimately influencing nonreciprocal transport properties. Our results quantitatively elucidate the role of in-plane geometric symmetry breaking in governing superconducting transport behavior.

## II. EXPERIMENT

The device patterns, with and without a triangular extension along the current channel, were fabricated on thermally oxidized silicon wafers using standard electron-beam lithography. A 40 nm thick niobium layer was then deposited onto the pre-patterned PMMA (950 A4) pits at room temperature via *d.c.* magnetron sputtering, with a base pressure below $4 \times 10^{-8}$ Torr. After lift-off in acetone, the final device structures were obtained. As shown in the scanning electron beam (SEM) image (**Fig. 1(a)**), the Hall bar channel had a fixed effective length of $100\ \mu\text{m}$ and terminal width of $5\ \mu\text{m}$. The constrained triangular extension of **Device #1** (base length $80\ \mu\text{m}$, height $30\ \mu\text{m}$) breaks the in-plane inversion symmetry orthogonal to the channel.

All cryogenic four-terminal measurements were conducted using a Quantum Design PPMS or a MultiFields Tech ColdTUBE, supplemented with external instruments. For *d.c.* measurements, as illustrated in **Fig. 1(b)**, a Keithley 6221 current source and a Keithley 2182A nanovoltmeter were used to measure the longitudinal resistance $R$. The $I_{\text{sd}}$-$V$ characteristics (**Fig. 1(c)**) were obtained with the 2182A triggered by the 6221, using a differential current sweep ($dI = 10\ \mu\text{A}$) to extract the corresponding differential voltage ($dV$). For nonlinear transport measurements, two Stanford Research SR830 lock-in amplifiers were employed to capture first- and second-harmonic responses to an *a.c.* sinusoidal current ($\tilde{I}_{\text{sd}},\ f = 133$ Hz), as shown in **Fig. 1(d)**. Further details on harmonic signal analysis are provided in **Supplemental Material S-1**.

## III. RESULTS AND DISCUSSION

**Figure 1(b)** presents the temperature-dependent resistance of **Device #1** under various out-of-plane magnetic fields ranging from 0 Oe to 10 kOe. To minimize thermal interference and ensure precise temperature control during measurements, the superconducting critical temperature was determined as $T_{c,0} = \frac{T(0.1R_N)+T(0.9R_N)}{2} = 6.0$ K during the heating process at zero magnetic field, where $R_{\text{N}}$ denotes the normal state resistance measured at 10 K. The observed degradation of superconductivity implies that defect-mediated mechanisms inhibit the formation of Cooper pairs[53]. The superconducting-to-normal transition at zero fields spans nearly 1 K, indicating a broad and gradual transition, which is likely due to localized electron scattering from defects or lattice distortions in the magnetron-sputtered Nb thin films, leading to spatial inhomogeneities in the superconducting order parameter[54]. As the applied magnetic field increases, quantized vortex cores are introduced, and their density rises[55]. Initially, these vortices remain pinned at defect sites and contribute minimally to dissipation. However, once thermal fluctuations or the Lorentz force exceed the pinning potential, the vortices become movable, leading to dynamic dissipation[56]. At 10 kOe, the transition extends beyond the low-temperature measurement limit, preventing the detection of superconducting zero resistance state.

**Figure 1(c)** displays the bias-direction-dependent $I_{sd}$-$V$ characteristics and the corresponding differential resistances measured at 5.1 K under 40 Oe. Two bidirectional current sweep measurements are conducted: the upward sequence ($-\mathrm{I_{max}} \rightarrow 0 \rightarrow +\mathrm{I_{max}}$) in the upper panel and the downward sequence ($+\mathrm{I_{max}} \rightarrow 0 \rightarrow -\mathrm{I_{max}}$) in the lower panel. The resulting $I_{sd}$-$V$ and differential resistance characteristics reveal distinct transport regimes, identified by different color annotations and labelled with Latin numerals I to V for clarity. In the upward sweep, hysteresis is observed for both positive and negative currents, corresponding to the phase transition between the superconducting and normal states. The transition points are defined as: retrapping current $I_r^-$, where device returns to the superconducting state, and superconducting critical current $I_c^+$, where the device transitions from superconducting to resistive state. Upon transitioning to the resistive state, a voltage step with a fixed resistance appears (region III to IV in **Fig 1(c)**), which may be attributed to the triangular extension of the device. This extension alters the internal distribution of pinning potentials, establishing a stable pinning platform that suppresses collective vortex motion phenomena (e.g., phase-slip lines) and compels vortices to propagate through individual trajectories. Such independent motion disrupts coordinated vortex dynamics while concurrently mitigating thermal fluctuation-induced vortex displacements. This dual mechanism ultimately manifests as abrupt voltage steps in the current-voltage characteristics, corresponding to discrete resistance states derived from quantized flux motion[37,57,58]. Subsequently, at region IV, a second abrupt resistance change occurs, followed by a smooth, arc-shaped $I$-$V$ curve extending from IV to V. This region reflects finite fluctuations in differential resistance, indicative of the gradual decoupling of superconducting pairs into diffusive quasiparticles[34,59]. As pair-breaking progresses, the supercurrent diminishes and ultimately vanishes[37,47,50,58]. The resulting resistance rise stabilizes at $R_N$, marking the completion of the superconducting-to-normal transition. Compared to the retrapping process, the transition at $I_c^+$ follows a "cold-to-hot" process, making it less susceptible to heat accumulation. Consequently, the following analysis mainly focuses on the devices' response to increasing input current from zero to a finite value.

**Figure 1(d)** presents first- and second-harmonic magnetoresistance performed at 2.4 K with varying *a.c.* currents $\tilde{I}_{sd}$. For currents below 100 μA, the second-harmonic signal remains negligible, indistinguishable from thermal noise. As the applied perpendicular magnetic field increases, superconductivity is progressively suppressed, with the upper critical field $\mu_0 H_{c,2}^{OOP} \approx 4$ T defined as field at which resistance reaches $0.5R_N$. When the *a.c.* bias current exceeds 400 μA, superconductivity is further weakened at fields lower than 4 T, broadening the superconducting-normal phase transition in the first-harmonic response. Analysis of the resistance slope $\mathrm{d}R/\mathrm{d}H^{OOP}$ reveals three distinct stages of flux penetration. Initially, as the magnetic field increases from 0 to approximately 8000 Oe, vortices rapidly penetrate weakly pinned regions, driving the system toward the normal state. As the field further increases, flux gradually penetrates stronger pinning regions, reducing the rate of resistance change. When the field exceeds $\mu_0 H_{c,2}^{OOP}$, superconductivity is entirely suppressed, and the sample transitions fully into the normal state. The emergence of a second-harmonic signal in the intermediate field range suggests the presence of asymmetric vortex transport, induced by the asymmetric pinning potential inherent to the device geometry. A more detailed analysis of this effect is provided in the subsequent sections and **Supplemental Material S-1**.

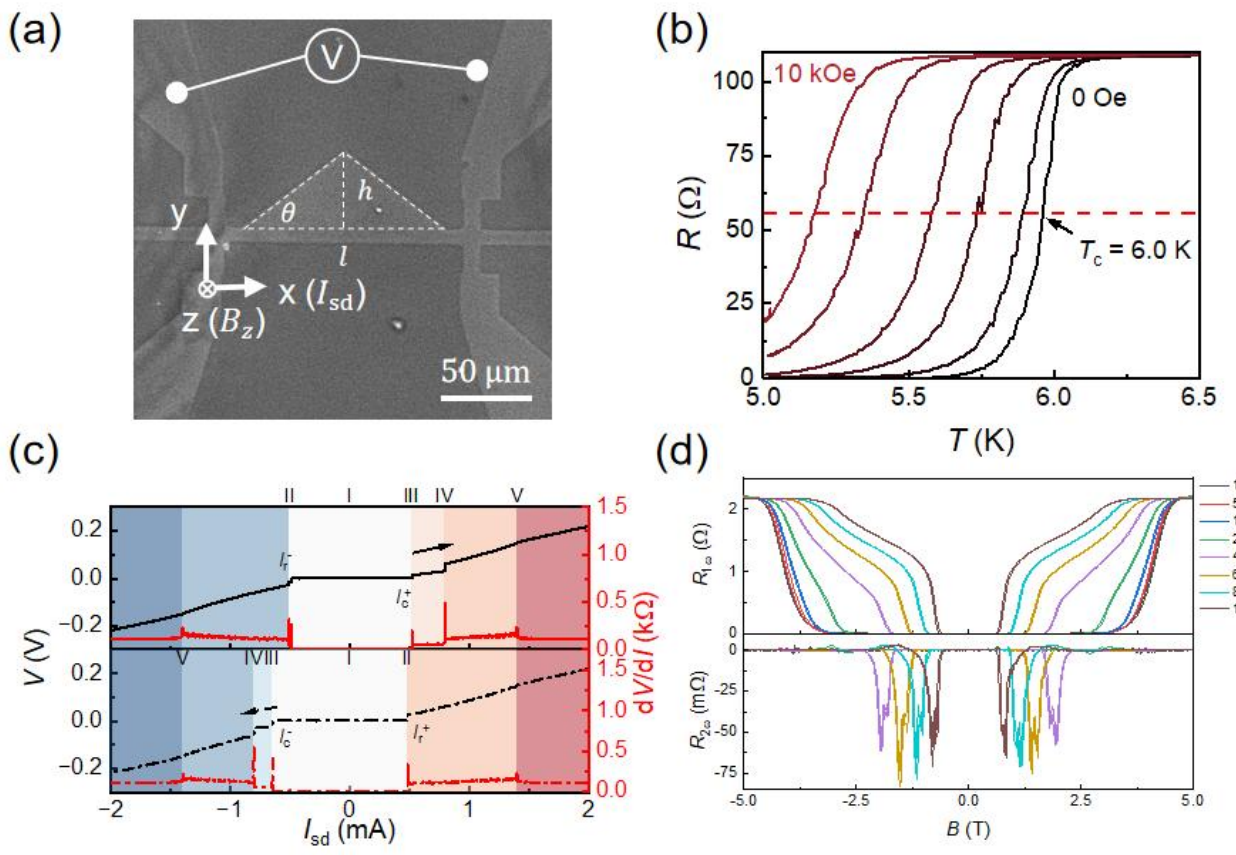

**FIG 1**. General characterization of the 40 nm thick Nb Hall bar device (**Device #1**). (a) SEM image and measurement configuration, with the white scale bar denoting 50 μm; (b) The temperature-dependent channel resistance $R_{xx}$ of the device under various magnetic fields ranging from 0 Oe to 10 kOe, with the data represented by a colour gradient from black to deep red. The light red dashed line indicates a resistance increase up to approximately 50% of the normal resistance in the absence of an external magnetic field. (c) Typical *I*-*V* characteristics and the corresponding differential resistance $dV/dI$ at 5.1 K under 40 Oe. The two panels, upper and lower, respectively show the current scanning from negative to positive (solid line, with the arrow showing the direction) and the scanning from positive to negative (dashed line, with the arrow showing the direction). The colour blocks indicate the regions dominated by different mechanisms, which are distinguished by the differential resistance. (d) Typical current density dependent first- and second-harmonic magnetoresistance at 2.4 K of 0.4 $T_c$.

To investigate the nonreciprocal transport properties of the device, we first examine its most fundamental manifestation: the superconducting diode effect. Analogous to the rectifying behavior in semiconductor p-n junctions, this phenomenon can be quantitatively characterized by analyzing the directional dependence of the critical current in the *I*-*V* characteristics. The asymmetry in critical current values for opposite current polarities serves as a direct indicator of nonreciprocal transport in the superconducting state. As shown in **Fig. 2(a)** and **(b)**, the two-dimensional color maps represent the longitudinal resistance ($V/I$) and the corresponding differential resistance ($dV/dI$), respectively. Based on the analysis from **Fig. 1(c)**, the evolution of the *I*-*V* characteristics under varying magnetic fields is clearly resolved. This enables the identification of distinct transport regimes and provides insight into the dominant mechanisms governing nonreciprocal behavior in the device.

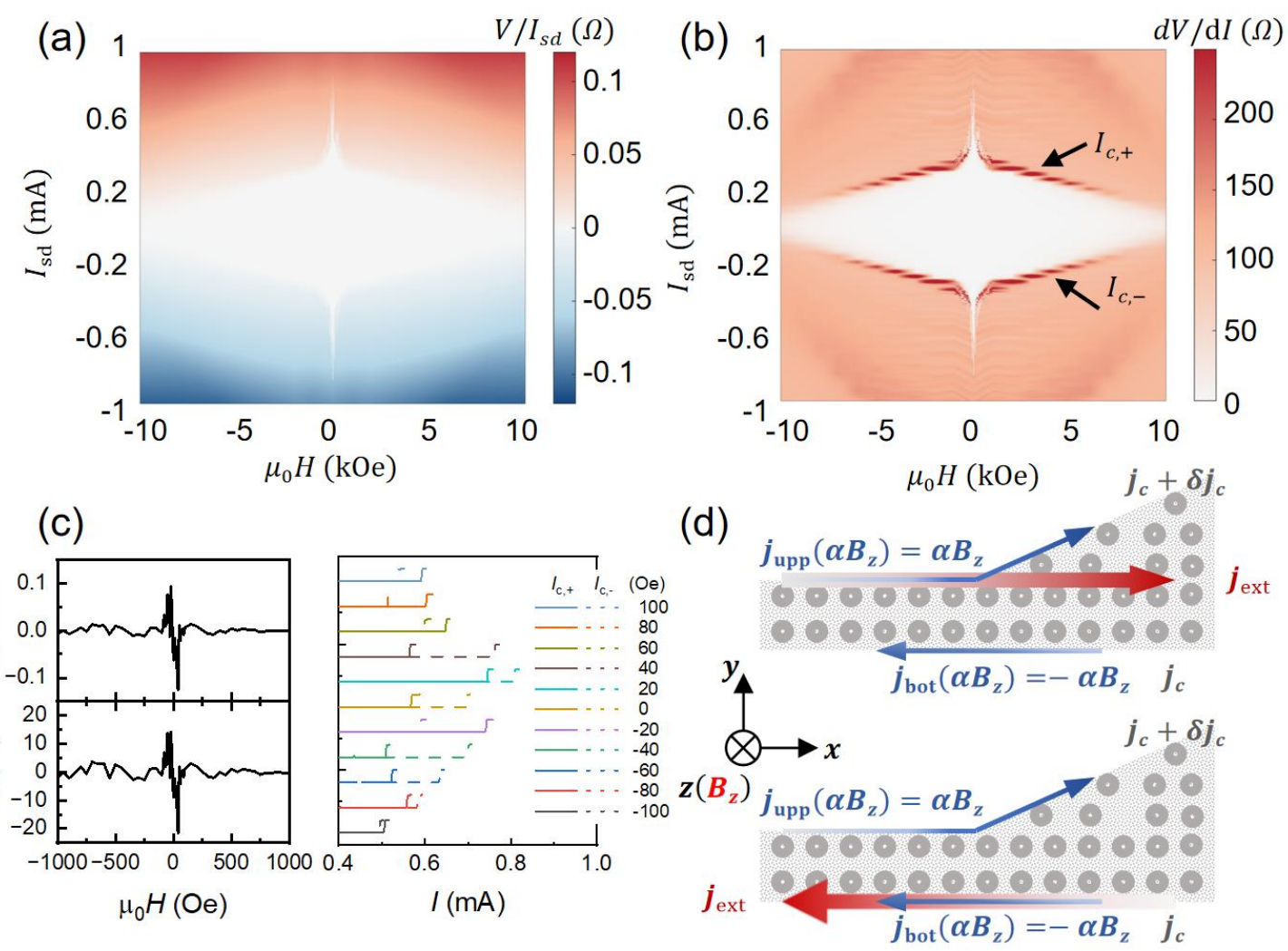

**FIG 2**. Nonreciprocal charge transport and superconducting diode effect in a 40 nm thick Nb Hall bar device (**Device #1**). (a) Two-dimensional colour map of the longitudinal resistance ($V/I$) as a function of source-drain current $I_{sd}$ and out-of-plane magnetic field $B$. (b) Corresponding differential resistance ($dV/dI$) mapping under the same parameter space. To minimize thermal fluctuations, analysis focuses exclusively on a unidirectional current sweep from zero to finite bias. (c) In the left panel, magnetic field (within range of $\pm 100$ Oe) dependence of nonreciprocal critical current difference $\Delta I_c = I_{c,+} - |I_{c,-}|$ and rectification efficiency Q(%). $|I_{c,\pm}|$ are extracted from the resistance and differential resistance characteristics in (a) and (b). In the right panel, selected IV characteristics within $\pm 40$ Oe, with the rotational mapping of negative-bias IV curves projected onto the positive current axis for comparative analysis. (d) Schematic illustration of the proposed mechanism for nonreciprocal transport. The triangular extension geometry induces an asymmetric distribution of superconducting critical conditions and magnetic flux vortex density between the extented ($j_c + \delta j_c$, $\delta j_c > 0$) and non-extended ($j_c$) regions. Under a positive out-of-plane magnetic field ($B_z > 0$), the Meissner screening currents exhibit opposite polarities, with magnitudes $\alpha B_z$ and $-\alpha B_z$ on the extented and non-extended sides, respectively, as indicated by the blue arrows. The red arrow denotes the external bias current ($j_{ext}$).

As the externally applied current increases from zero to a finite value, the critical current ($I_{c,\pm}$)—at which the device transitions from the superconducting to the resistive state—monotonically decreases with increasing magnetic field. This reduction in $|I_{c,\pm}|$ arises from the field-induced suppression of superconductivity, which facilitates the penetration of magnetic flux vortices into the superconducting material. As a result, a lower applied current is sufficient to initiate vortex motion, reducing the critical current threshold. For magnetic fields up to 1000 Oe, the vortex density remains relatively low, and vortices are predominantly confined by weak pinning potentials. In this regime, a relatively high current density is required to trigger the superconducting-to-resistive transition. As the applied current increases, the device remains in the vortex creep regime, where a limited number of vortices undergo thermally activated motion under the influence of the weak Lorentz force. This behavior manifests in the *I-V* characteristics as a stable resistive state[47,48,60]. Only when the current exceeds a second critical threshold does the Lorentz force surpass the pinning potential, leading to complete vortex depinning, the breakdown of superconductivity, and a transition to the normal state. The critical currents ($|I_{c,\pm}|$) at different magnetic fields can be systematically extracted from the longitudinal resistance and differential resistance mappings in **Fig. 2(a)** and **(b)**, with summarized results provided in **FIG S2**, **Supplementary Note S-3**. Specifically, when the applied current lies between $I_{c,+}$ and $|I_{c,-}|$, the device exhibits directionally selective transport: superconductivity is preserved for one current polarity ($I_{bias}$), while the opposite polarity ($-I_{bias}$) induces a transition to the resistive state, thereby manifesting the superconducting diode effect.

To quantify this effect, we analyze the magnetic field dependence of the critical current asymmetry ($\Delta I_c = I_{c,+} - |I_{c,-}|$) and the corresponding rectification efficiency ($Q = 2\Delta I_c/(I_{c,+} + |I_{c,-}|)$). These parameters, crucial for evaluating superconducting diode performance, are systematically illustrated in the left panel of **Fig. 2(c)**. Under an applied magnetic field of ±40 Oe, the diode device demonstrates an optimal operational current window centered around 0.1 mA, where the rectification efficiency reaches its maximum value of approximately 20%. **Fig. 2(d)** further elucidates the transport mechanisms governing nonreciprocal behavior in the presence of geometric symmetry breaking. The triangular extension and structural defects along the edges induce an imbalance in critical conditions between the upper ($j_c + \delta j_c$, where $\delta j_c > 0$ due to the extension) and lower ($j_c$, where the extension is absent) channel boundaries. Under tan applied out-of-plane magnetic field ($B_z$), Meissner currents develop at both edges—$+\alpha B_z$ at the upper edge and $-\alpha B_z$ at the lower edge—depicted by the blue arrows in the figure. In contrast to prior studies[19,25], the intentionally engineered triangular extension significantly amplifies differences in superconducting critical parameters ($\delta j_c$) along the channel boundaries, facilitating an anisotropic transport response.

The triangular extension also plays a crucial role in modulating vortex density distribution, as illustrated in **Fig. 2(d)**. This effect is particularly pronounced in the angle region between the extension and the adjacent channel edge, where vortex dynamics are influenced by a locally weak pinning potential[25]. Under an applied $B_z$ field, the region near the triangular extension exhibits lower vortex density and increased inter-vortex spacing, making individual vortices more susceptible to localized weak pinning effects. In contrast, the region without the extension exhibits higher vortex density, where stronger vortex-vortex interactions increase the critical depinning force required for vortex motion. As shown in **Fig. 2(d)**, these spatial variations in vortex dynamics lead to directionally dependent vortex motion. Under forward current bias (upper panel), the Lorentz force preferentially drives vortices along the extended region, where the reduced pinning potential facilitates a pronounced resistance transition upon vortex depinning. Conversely, under reverse bias (lower panel), vortices are pushed toward the high-potential region, leading to vortex depletion in the extended region and requiring higher current densities for thermal activation and vortex propulsion. Above analysis provides a phenomenological interpretation of nonreciprocal transport phenomena, linking geometric asymmetry, vortex dynamics, and superconducting diode behavior.

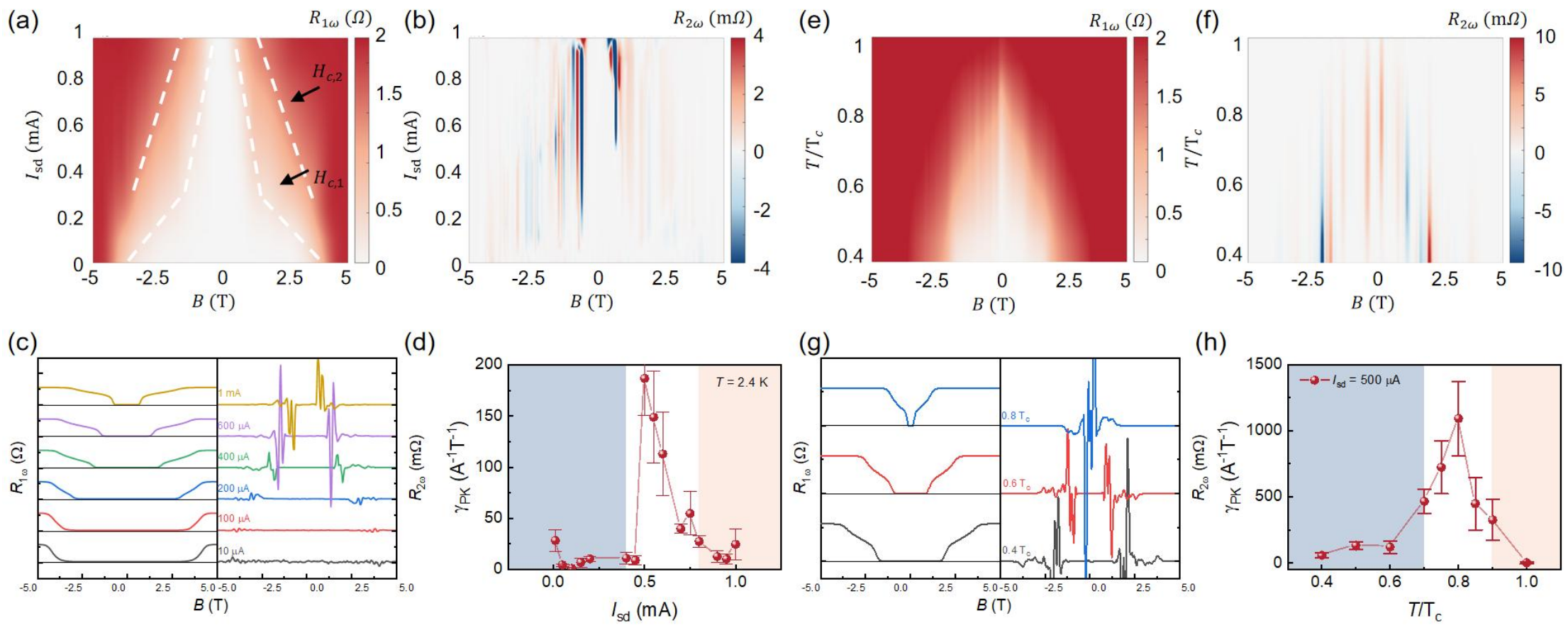

**FIG 3**. Nonreciprocal charge transport and magnetochiral anisotropy in a 40 nm thick Nb Hall bar device (**Device #1**). (a) (b) Two-dimensional colour map of the first-harmonic ($V_{1\omega}/I_{sd}$) and second-harmonic ($V_{2\omega}/I_{sd}$) magnetoresistance as a function of source-drain current $I_{sd}$ and out-of-plane magnetic field $B$ at $T = 0.4\,T_c$. The white dashed line in (a) schematically indicates the approximate position of the critical field corresponding to the resistive transition. (c) Selective magnetic field dependence of the first-harmonic resistance ($V_{1\omega}/I_{sd}$, scaled to 4 Ω) and the second-harmonic resistance ($V_{2\omega}/I_{sd}$, scaled to 4 mΩ), as extracted from (a) and (b), with an applied vertical shift for clarity. (d) Current density dependence of $\gamma$ obtained from (a) and (b). (e) (f) Two-dimensional colour map of the first-harmonic ($V_{1\omega}/I_{sd}$) and second-harmonic ($V_{2\omega}/I_{sd}$) resistance as a function of normalized temperature $T/T_c$ and out-of-plane magnetic field $B$ with the bias source-drain current $I_{sd} = 500$ μA. (g) Selective magnetic field dependence of the first-harmonic resistance ($V_{1\omega}/I_{sd}$, scaled to 4 Ω) and the second-harmonic resistance ($V_{2\omega}/I_{sd}$, scaled to 10 mΩ), as extracted from (e) and (f), with an applied vertical shift for clarity. (h) Temperature dependence of $\gamma$ obtained from (e) and (f). Error bars are derived from the uncertainty associated with multiple peaks in second- harmonic resistance analysis. To minimize thermal fluctuations, all experimental analyses were focused on the magnetic field sweep trajectories from zero to finite values.

To further investigate the nonreciprocal transport behavior of the device, we analyze it from the perspective of nonlinear response. Previous theoretical and experimental studies have demonstrated that in noncentrosymmetric semiconductor systems with broken inversion symmetry, a nonlinear second-harmonic term $I \cdot B$ persists in the *a.c.* voltage response when time-reversal symmetry is simultaneously broken by an external magnetic field[6,43,61,62]. We hereby extend this methodology to superconducting systems. Theoretical formulations in analyzing nonlinear responses of noncentrosymmetric structures often prove inadequate for distinguishing between different physical contributions to the experimentally observed magnetoresistance behavior[9]. As evidenced in **Figure 1(d)**, the raw magnetoresistance exhibit pronounced even-symmetric components with respect to the magnetic field, which may originate from ordinary magnetoresistance, nonlinear conductivity (NLC), or other unidentified mechanisms in this system. The potential existence of NLC components in this architecture remains unresolved and constitutes an active research domain in nonlinear transport phenomena of noncentrosymmetric materials. For our primary focus on magnetic chiral anisotropy analysis, we have performed systematic antisymmetrization processing of the raw data to isolate the relevant second-harmonic component $I \cdot B$, details can be found in **Supplementary Materials S-1**. The post-processed alternating response $R_{2\omega}$ of the device exhibits a dependence on both $I$ and $B$, signifying the presence of magnetochiral anisotropy. Consequently, the existence of the coefficient $\gamma$ provides direct evidence of broken inversion symmetry within the device structure. In our experimental design, $\gamma$ serves as a quantitative measure of how the extension structures influence nonreciprocal transport properties.

At 0.4 $T_{c,0}$, systematic measurements of harmonic magnetoresistance were conducted under a forward-bias sinusoidal *a.c.* current $\tilde{I}_{sd}$, while sweeping the magnetic field in the sequence -5 → 0 → 5 → 0 → -5 T. The acquired data were subsequently processed into two-dimensional color maps, enabling a comprehensive analysis of magnetochiral anisotropy. To eliminate potential artifacts arising from thermal hysteresis effects, the experimental analysis specifically focuses on field sweep from zero to finite values, as illustrated in **Fig. 3(a)** and **(b)**. As demonstrated in **Fig 1(d)**, the magnetoresistance transition exhibits distinct regimes, governed by mechanisms demarcated by two critical magnetic fields $H_{c,1}$ and $H_{c,2}$ as the bias current increases. A similar trend is observed in **Fig 3(a)**, where these transitional boundaries are schematically delineated by white dashed lines.

**Figure 3(c)** presents $R_{1\omega}$ and $R_{2\omega}$ as functions of magnetic field under different current biases. The magnetic field values corresponding to the peak positions in the second-harmonic signal exhibit a direct correlation with magnetoresistance variations in the first-harmonic signal, specifically aligning with the superconducting-to-normal phase transition at $H_{c,1}$. This correlation reflects underlying vortex dynamics. At lower bias currents, the applied current is insufficient to drive the collective motion of high-density vortices induced by larger critical magnetic fields, resulting in a diminished second-harmonic signal. As the bias current increases, the critical magnetic field progressively decreases. When the applied current reaches a threshold sufficient to mobilize individual vortices constrained by weak pinning potentials, a significant enhancement in the second-harmonic signal is observed. According to **Eq. S-3** and **Eq. S-4** from **Supplementary Note S-1**, the magnetochiral anisotropy coefficient $\gamma$ is given by $\gamma = \frac{\sqrt{2}V_{2\omega}}{V_{1\omega}BI}$, where $V_{2\omega}$ represents the peak amplitude of the second-harmonic response, while $V_{1\omega}$ and $B$ denote the corresponding first-harmonic response and magnetic field, respectively (further details in **Supplementary Note S-2**). The relationship between $\gamma$ and the effective bias current $I_{\rm sd}$ is illustrated in **Fig. 3(d)**. It is crucial to note that pronounced fluctuations in second-harmonic responses and broadening of the peak bandwidth introduce uncertainty in determining the exact positions of $R_{2\omega}$. This intrinsic variability necessitates the incorporation of error bars in subsequent quantitative analyses. A similar experimental approach was used to analyze the normalized temperature dependence of $\gamma$ with $I_{\rm sd} = 500$ μA, as shown in **Fig. 3(e)–(h)**. The variations in $\gamma$ as a function of bias current (**Fig. 3(d)**) and temperature (**Fig. 3(h)**) indicate that the vortex-mediated nonreciprocal transport mechanism requires the device to operate below the critical threshold. However, when the device is deep in the superconducting condensed state, at lower current or temperature, it remains sufficiently "cold" and necessitates a stronger magnetic field to suppress superconductivity. In this regime, the magnetic field overcomes the weak pinning potential, increasing vortex density and intensifying vortex-vortex interactions, which leads to collective vortex dynamics. Consequently, the manifestation of nonreciprocal transport becomes less prominent. As the bias current or temperature increases, $\gamma$ reaches a peak, corresponding to the regime where vortices are most active. Beyond this point, as the bias current or temperature approaches the critical conditions, superconductivity is gradually suppressed due to enhanced fluctuations and dissipation. Consequently, vortices depin and transition into quasiparticles, leading to a decrease in γ at this stage. Beyond this phenomenological interpretation, further theoretical analysis is required to fully elucidate the underlying physical mechanisms.

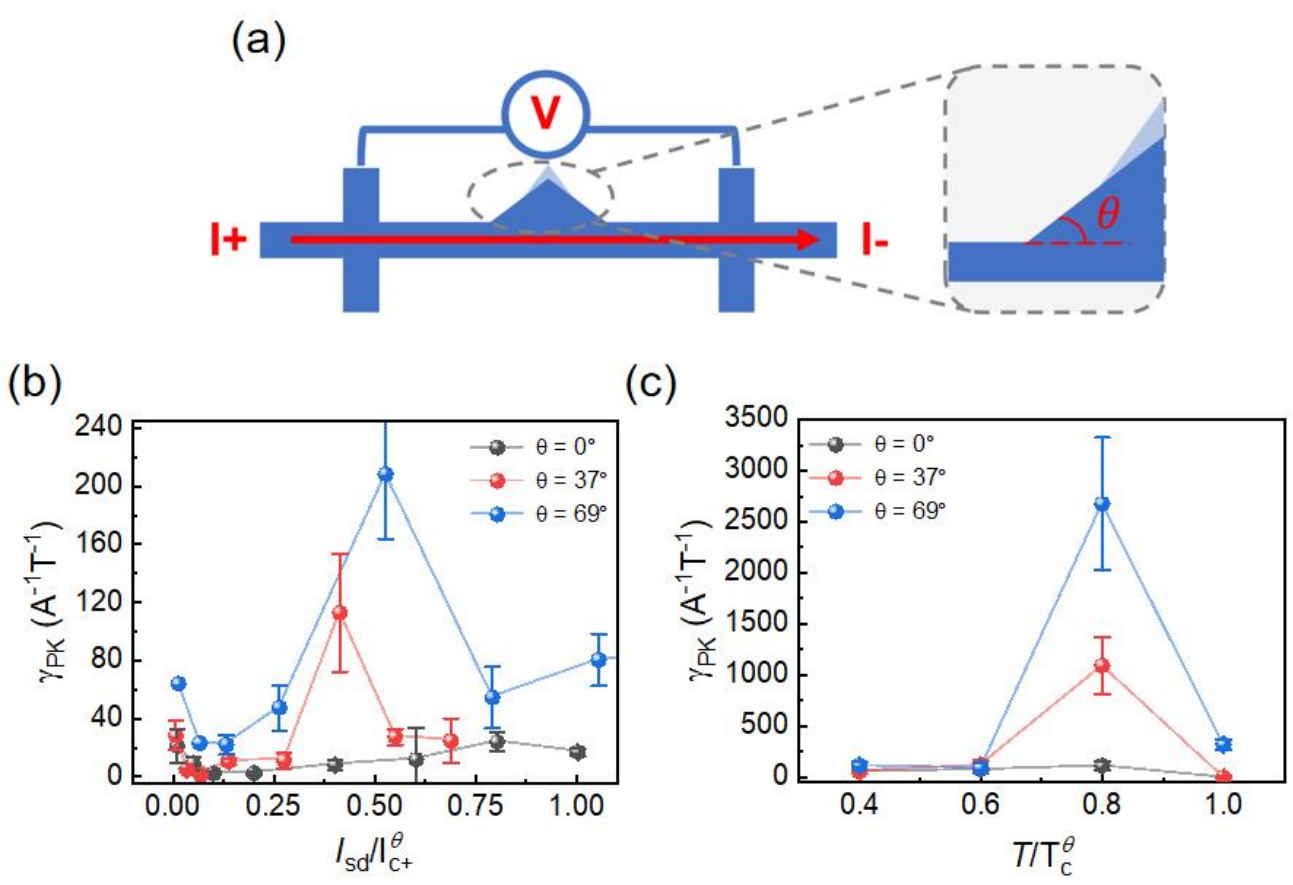

**FIG 4**. Quantitative comparison of magnetochiral anisotropy and extension-induced symmetry breaking. (a) Schematic illustration of the Hall bar device structures. A zoom-in view of one channel

edge highlights the parameter $\theta$, which defines the angular orientation between the triangular extension feature and the adjacent channel edge, serving as the key distinguishing factor among the device variations. The study examines three distinct structural configurations: $\theta$ (in degrees) = 0° corresponds to a conventional symmetric Hall bar; $\theta = 37°$, associated with a base width of 80 μm and a vertical height of 30 μm; $\theta = 69°$, characterized by a reduced base width of 30 μm and an increased vertical height of 40 μm. (b) Normalized current density dependence of $\gamma$ at $T = 0.4\,\mathrm{T}_{\mathrm{c,0}}^{\theta}$. (c) Normalized temperature dependence of $\gamma$ with the bias source-drain current $I_{\mathrm{sd}} = 500$ μA. Both (b) and (c) demonstrate that the introduction of the triangular extension feature disrupts in-plane symmetry, leading to an enhancement of the magnetochiral anisotropy coefficient.

We further undertake a quantitative analysis of the influence of extension on nonreciprocal transport. As is shown in **Fig. 4(a)**, in addition to **Device #1**, which features a triangular extension with a base of $l = 80$ μm and a height of $h = 30$ μm, two control devices were designed: a symmetric Hall bar without a triangular extension (**Device #2**) and a Hall bar with a triangular extension of $l = 30$ μm base and $h = 40$ μm height (**Device #3**). The vortex density distribution, which predominantly governs the nonreciprocal transport as evidenced by prior study on vortex-related crowding effects in superconducting nanowire diode[25], necessitates the definition of a key parameter corresponding to the local region at the angle between the extended feature and the adjacent channel edge. **Figure 4(b)** presents the normalized bias current $I_{\mathrm{sd}}/\mathrm{I}_{\mathrm{c+}}^{\theta}$ dependence of $\gamma$ for the three devices at $0.4\,\mathrm{T}_{\mathrm{c,0}}^{\theta}$. In this analysis, $\mathrm{I}_{\mathrm{c+}}^{\theta}$ and $\mathrm{T}_{\mathrm{c,0}}^{\theta}$ denote the superconducting critical parameters for different devices $\theta$. Since we are exclusively considering the positive current branch in the MChA analysis, the value of $\mathrm{I}_{\mathrm{c+}}^{\theta}$ is determined at the first critical current value where the system transitions from the superconducting state to the resistive state during the current sweep from $-\mathrm{I}_{\max}$ to $+\mathrm{I}_{\max}$.

Compared to the symmetric Hall bar ( $\theta = 0$ ), the introduction of a triangular extension enhances the $\gamma$ values, further confirming the role of inversion symmetry breaking. Moreover, a larger $\theta$ angle corresponds to a higher peak value of $\gamma$. **Figure 4(c)** shows the normalized temperature dependence of $\gamma$ at a bias current of 500 μA, demonstrating that the inclusion of the triangular extension structure similarly amplifies $\gamma$ across different temperature conditions. Through normalized analysis, it is recognized that all three devices exhibit peak performance near 0.8 $\mathrm{T}_{\mathrm{c,0}}^{\theta}$, suggesting that the variation in the $\gamma$ is closely associated with its critical value. This correlation is also evident in **Figure 4(b)**, where peak performance occurs around 0.4 $\mathrm{I}_{\mathrm{c+}}^{\theta}$. It is noteworthy that the terminal regions share identical cross-sectional areas, ensuring uniform current density at the terminals under the same applied bias current, differences in the triangular extension structure influence the carrier density distribution as well as local superconductivity within the device. A larger triangular extension area strengthens superconductivity, leading to a higher critical current. Consequently, **Device #3**, which has a relatively smaller triangular extension area of $60\ \mu m^2$, exhibits a lower critical current compared to **Device #1** with a larger area of $120\ \mu m^2$.

## IV. CONCLUSION

In this study, we systematically investigated the influence of geometric symmetry modulation on nonreciprocal transport in superconducting devices using tailored triangular extended Hall bar structures. By comparing devices with asymmetric triangular extensions to fully symmetric controls, we demonstrate that engineered in-plane geometric symmetry breaking significantly enhances the superconducting diode effect, as quantified by MChA coefficient $\gamma$ extracted from second-harmonic nonlinear magnetoresistance measurements. A key finding is the positive correlation between the nonreciprocal coefficient and the base angle of the triangular extension, underscoring the crucial role of geometric design in tuning vortex dynamics and spatial inversion symmetry breaking under time-reversal symmetry-breaking conditions. Furthermore, our results establish second-harmonic analysis as a powerful diagnostic tool for disentangling intrinsic symmetry-breaking mechanisms from extrinsic effects such as flux vortex pinning by defects or inhomogeneous current distributions. The observed interplay between mesoscopic geometry, vortex motion, and macroscopic nonreciprocity provides a guiding framework for designing superconducting rectification devices and engineering cryogenic quantum systems with tailored transport properties. Looking ahead, this paradigm could be extended to other symmetry-broken geometries, hybrid material systems, or topological superconductors, where competing symmetry-breaking mechanisms may yield even richer nonlinear responses.

## ACKNOWLEDGEMENTS

This work was supported by the National Key Research and Development Program of China (Grant No. 2022YFA1402604), the Natural Science Foundation of China (52201200), and Research Funds for the Central Universities.

## REFERENCE

[1] P. J. W. Moll, V. B. Geshkenbein, *Nature Phys.* **2023**, *19*, 1379.
[2] M. Nadeem, M. S. Fuhrer, X. Wang, *Nat. Rev. Phys.* **2023**, *5*, 558.
[3] Y. He, J. Li, Q. Wang, H. Matsuki, G. Yang, *Adv. Devices Instrum.* **2023**, *4*, 0035.
[4] D. Li, Z. Lu, W. Cheng, X. Shi, L. Hu, X. Ma, Y. Liu, Y. M. Itahashi, T. Shitaokoshi, P. Li, H. Zhang, Z. Liu, F. Qu, J. Shen, Q. Chen, K. Jin, J. Cheng, J. Hänisch, H. Yang, G. Liu, L. Lu, X. Dong, Y. Iwasa, J. Hu, T. Kokkeler, I. Tokatly, F. S. Bergeret, N. Nagaosa, Y. Yanase, *Annu. Rev. Condens. Matter Phys.* **2024**, *15*, 63.
[5] Y. Cheng, Q. Shu, H. He, B. Dai, K. L. Wang, *Adv. Mater.* **2025**, 2415480.
[6] G. L. J. A. Rikken, J. Fölling, P. Wyder, *Phys. Rev. Lett.* **2001**, *87*, 236602.
[7] R. Wakatsuki, Y. Saito, S. Hoshino, Y. M. Itahashi, T. Ideue, M. Ezawa, Y. Iwasa, N. Nagaosa, *Sci. Adv.* **2017**, *3*, e1602390.
[8] Y. Tokura, N. Nagaosa, *Nat. Commun.* **2018**, *9*, 3740.
[9] M. Suárez-Rodríguez, F. De Juan, I. Souza, M. Gobbi, F. Casanova, L. E. Hueso, *Non-linear Transport in Non-centrosymmetric Systems: From Fundamentals to Applications*, arXiv, **2024**.
[10] F. Ando, Y. Miyasaka, T. Li, J. Ishizuka, T. Arakawa, Y. Shiota, T. Moriyama, Y. Yanase, T. Ono, *Nature* **2020**, *584*, 373.
[11] S. Ilić, F. S. Bergeret, *Phys. Rev. Lett.* **2022**, *128*, 177001.
[12] R. Cai, D. Yue, W. Qiao, L. Guo, Z. Chen, X. C. Xie, X. Jin, W. Han, *Phys. Rev. B* **2023**, *108*, 064501.
[13] S. Ilić, P. Virtanen, D. Crawford, T. T. Heikkilä, F. S. Bergeret, *Phys. Rev. B* **2024**, *110*, L140501.
[14] H. Huang, T. de Picoli, J. I. Väyrynen, *Appl. Phys. Lett.* **2024**, *125*, 032602.
[15] Y. Mao, Q. Yan, Y.-C. Zhuang, Q.-F. Sun, *Phys. Rev. Lett.* **2024**, *132*, 216001.
[16] H. Narita, J. Ishizuka, D. Kan, Y. Shimakawa, Y. Yanase, T. Ono, *Adv. Mater.* **2023**, *35*, 2304083.
[17] N. F. Q. Yuan, L. Fu, *Proc. Natl. Acad. Sci. U.S.A.* **2022**, *119*, e2119548119.
[18] B. Pal, A. Chakraborty, P. K. Sivakumar, M. Davydova, A. K. Gopi, A. K. Pandeya, J. A. Krieger, Y. Zhang, M. Date, S. Ju, N. Yuan, N. B. M. Schröter, L. Fu, S. S. P. Parkin, *Nature Phys.* **2022**, *18*, 1228.
[19] Y. Hou, F. Nichele, H. Chi, A. Lodesani, Y. Wu, M. F. Ritter, D. Z. Haxell, M. Davydova, S. Ilić, O. Glezakou-Elbert, A. Varambally, F. S. Bergeret, A. Kamra, L. Fu, P. A. Lee, J. S. Moodera, *Phys. Rev. Lett.* **2023**, *131*, 027001.
[20] C.-S. Lee, B. Jankó, I. Derényi, A.-L. Barabási, *Nature* **1999**, *400*, 337.
[21] E. Zhang, X. Xu, Y.-C. Zou, L. Ai, X. Dong, C. Huang, P. Leng, S. Liu, Y. Zhang, Z. Jia, X. Peng, M. Zhao, Y. Yang, Z. Li, H. Guo, S. J. Haigh, N. Nagaosa, J. Shen, F. Xiu, *Nat. Commun.* **2020**, *11*, 5634.
[22] M. Castellani, O. Medeiros, A. Buzzi, R. A. Foster, M. Colangelo, K. K. Berggren, *A superconducting full-wave bridge rectifier*, arXiv, **2025**.
[23] Y. Togawa, K. Harada, T. Akashi, H. Kasai, T. Matsuda, F. Nori, A. Maeda, A. Tonomura, *Phys. Rev. Lett.* **2005**, *95*, 087002.
[24] S. S. Ustavschikov, M. Yu. Levichev, I. Yu. Pashenkin, N. S. Gusev, S. A. Gusev, D. Yu. Vodolazov, *J. Exp. Theor. Phys.* **2022**, *135*, 226.
[25] X. Zhang, Q. Huan, R. Ma, X. Zhang, J. Huang, X. Liu, W. Peng, H. Li, Z. Wang, X. Xie, L. You, *Adv. Quantum Technol.* **2024**, *7*, 2300378.
[26] L. R. Cadorim, E. Sardella, C. C. de S. Silva, *Phys. Rev. Applied* **2024**, *21*, 054040.
[27] K. Yasuda, H. Yasuda, T. Liang, R. Yoshimi, A. Tsukazaki, K. S. Takahashi, N. Nagaosa, M. Kawasaki, Y. Tokura, *Nat. Commun.* **2019**, *10*, 2734.

[28] J. Jo, Y. Peisen, H. Yang, S. Mañas-Valero, J. J. Baldoví, Y. Lu, E. Coronado, F. Casanova, F. S. Bergeret, M. Gobbi, L. E. Hueso, *Nat. Commun.* **2023**, *14*, 7253.
[29] A. Gutfreund, H. Matsuki, V. Plastovets, A. Noah, L. Gorzawski, N. Fridman, G. Yang, A. Buzdin, O. Millo, J. W. A. Robinson, Y. Anahory, *Nat. Commun.* **2023**, *14*, 1630.
[30] S. Sengupta, M. Monteverde, S. Loucif, F. Pallier, L. Dumoulin, C. Marrache-Kikuchi, *Phys. Rev. B* **2024**, *109*, L060503.
[31] S. Sengupta, A. Farhadizadeh, J. Youssef, S. Loucif, F. Pallier, L. Dumoulin, K. Saha, S. Pujari, M. Oden, C. Marrache-Kikuchi, M. Monteverde, *Transverse resistance due to electronic inhomogeneities in superconductors*, arXiv, **2024**.
[32] Du W.-S., Chen W., Zhou Y., Zhou T., Liu G., Xiao Z., Zhang Z., Miao Z., Jia H., Liu S., Zhao Y., Zhang Z., Chen T., Wang N., Huang W., Tan Z.-B., Chen J.-J., Yu D.-P., *Phys. Rev. B* **2024**, *110*, 174509.
[33] D. Suri, A. Kamra, T. N. G. Meier, M. Kronseder, W. Belzig, C. H. Back, C. Strunk, *Appl. Phys. Lett.* **2022**, *121*, 102601.
[34] N. Satchell, P. Shepley, M. Rosamond, G. Burnell, *J. Appl. Phys.* **2023**, *133*, 203901.
[35] B. L. T. Plourde, D. J. Van Harlingen, D. Yu. Vodolazov, R. Besseling, M. B. S. Hesselberth, P. H. Kes, *Phys. Rev. B* **2001**, *64*, 014503.
[36] D. Y. Vodolazov, F. M. Peeters, *Phys. Rev. B* **2005**, *72*, 172508.
[37] G. R. Berdiyorov, A. K. Elmurodov, F. M. Peeters, D. Y. Vodolazov, *Phys. Rev. B* **2011**, *79*, 174506.
[38] S. Hoshino, R. Wakatsuki, K. Hamamoto, N. Nagaosa, *Phys. Rev. B* **2018**, *98*, 054510.
[39] C. Chen, Y. Liu, Y. Chen, Y. N. Hu, T. Z. Zhang, D. Li, X. Wang, C. X. Wang, Z. Y. W. Lu, Y. H. Zhang, Q. L. Zhang, X. L. Dong, R. Wang, D. L. Feng, T. Zhang, *Phys. Rev. X* **2024**, *14*.
[40] Y. M. Itahashi, T. Ideue, S. Hoshino, C. Goto, H. Namiki, T. Sasagawa, Y. Iwasa, *Nat. Commun.* **2022**, *13*, 1659.
[41] J. Ma, R. Zhan, X. Lin, *Adv. Phys. Res.* **2025**, 2400180.
[42] Wang C., Hu G., Ma X., Tan H., Wu J., Feng Y., Wang S., Li R., Zheng B., He J. J., Xiang B., *Phys. Rev. Appl.* **2024**, *22*, 064017.
[43] T. Ideue, K. Hamamoto, S. Koshikawa, M. Ezawa, S. Shimizu, Y. Kaneko, Y. Tokura, N. Nagaosa, Y. Iwasa, *Nature Phys.* **2017**, *13*, 578.
[44] H. Wu, Y. Wang, Y. Xu, P. K. Sivakumar, C. Pasco, U. Filippozzi, S. S. P. Parkin, Y.-J. Zeng, T. McQueen, M. N. Ali, *Nature* **2022**, *604*, 653.
[45] Y. M. Itahashi, T. Ideue, Y. Saito, S. Shimizu, T. Ouchi, T. Nojima, Y. Iwasa, *Sci. Adv.* **2020**, *6*, eaay9120.
[46] Y. Wu, Q. Wang, X. Zhou, J. Wang, P. Dong, J. He, Y. Ding, B. Teng, Y. Zhang, Y. Li, C. Zhao, H. Zhang, J. Liu, Y. Qi, K. Watanabe, T. Taniguchi, J. Li, *npj Quantum Mater.* **2022**, *7*, 105.
[47] S. Bhattacharya, M. J. Higgins, *Phys. Rev. Lett.* **1993**, *70*, 2617.
[48] M. Chandran, R. T. Scalettar, G. T. Zimányi, *Phys. Rev. B* **2003**, *67*, 052507.
[49] D. Y. Vodolazov, F. M. Peeters, *Phys. Rev. B* **2007**, *76*, 014521.
[50] O. V. Dobrovolskiy, D. Y. Vodolazov, F. Porrati, R. Sachser, V. M. Bevz, M. Y. Mikhailov, A. V. Chumak, M. Huth, *Nat. Commun.* **2020**, *11*, 3291.
[51] S. Kumar Ojha, P. Mandal, S. Kumar, J. Maity, S. Middey, *Commun. Phys.* **2023**, *6*, 126.
[52] J. J. He, Y. Tanaka, N. Nagaosa, *New J. Phys.* **2022**, *24*, 053014.
[53] D. Bromley, A. J. Wright, L. A. H. Jones, J. E. N. Swallow, T. Beesley, R. Batty, R. S. Weatherup, V. R. Dhanak, L. O'Brien, *Sci. Rep.* **2022**, *12*, 7786.

[54] J. Bardeen, M. J. Stephen, *Phys. Rev.* **1965**, *140*, A1197.
[55] J.-Y. Ge, J. Gutierrez, V. N. Gladilin, J. T. Devreese, V. V. Moshchalkov, *Nat. Commun.* **2015**, *6*, 6573.
[56] L. He, J. Wang, *Nanotechnology* **2011**, *22*, 445704.
[57] A. G. Sivakov, A. M. Glukhov, A. N. Omelyanchouk, Y. Koval, P. Müller, A. V. Ustinov, *Phys. Rev. Lett.* **2003**, *91*, 267001.
[58] G. Berdiyorov, K. Harrabi, J. P. Maneval, F. M. Peeters, *Supercond. Sci. Technol.* **2015**, *28*, 025004.
[59] E. Zhang, J. Zhi, Y.-C. Zou, Z. Ye, L. Ai, J. Shi, C. Huang, S. Liu, Z. Lin, X. Zheng, N. Kang, H. Xu, W. Wang, L. He, J. Zou, J. Liu, Z. Mao, F. Xiu, *Nat. Commun.* **2018**, *9*, 4656.
[60] S. Ryu, M. Hellerqvist, S. Doniach, A. Kapitulnik, D. Stroud, *Phys. Rev. Lett.* **1996**, *77*, 5114.
[61] M. Akaike, Y. Nii, H. Masuda, Y. Onose, *Phys. Rev. B* **2021**, *103*, 184428.
[62] T. Ideue, Y. Iwasa, *Annu. Rev. Condens. Matter Phys.* **2021**, *12*, 201.